\documentclass[aps,showpacs,amsfonts,latexsym,amsthm]{revtex4}

\usepackage{latexsym}
\usepackage{graphicx}
\newtheorem{theorem}{Theorem}
\newtheorem{lemma}[theorem]{Lemma}
\newtheorem{proposition}[theorem]{Proposition}

\newenvironment{proof}[1]{%
  \begin{trivlist}{}{\setlength{\topsep}{0cm}\setlength{\partopsep}{0cm}}
  \item \textbf{#1.\@}\hspace*{1ex}\ignorespaces}%
  {\makebox[0cm]{}\nolinebreak\hfill$\Box$\end{trivlist}}

\def\ket#1{{|{#1}\rangle}}
\def\bra#1{{\langle{#1}|}}
\def\T{{\dagger}}
\def\imagi{{\imath}}
\def\del{\mbox{\rm del}}
\def\norm#1{{\left\|{#1}\right\|}}
\def\glb1q{\diamond}

\def\Z{{\mathbb Z}}

\def\R{{\mathbb R}}
\def\C{{\mathbb C}}

\begin{document}

\title{Quantum Computing on Lattices using Global Two-Qubit Gates}
\author{G. Ivanyos}
\email{Gabor.Ivanyos@sztaki.hu}
\affiliation{
Computer and Automation Research Institute,
Hungarian Academy of Sciences,
\\
{H-1518 Budapest, P.O. Box 63., Hungary.}
} 
\author{S. Massar}
\email{smassar@ulb.ac.be}
\affiliation{Laboratoire d'Information Quantique and QUIC, {C.P.}
165/59, Av. F. D. Roosevelt 50, B-1050 Bruxelles, Belgium} 
\author{A. B. Nagy}
\email{nagy@math.bme.hu}
\affiliation{Budapest University of Technology and Economics,
\\
H-1521 Budapest, P.O. Box 91., Hungary.}  

\begin{abstract}
We study the computation power of lattices composed of two dimensional
systems (qubits) on which translationally invariant  
global two-qubit gates can
be performed. We show that if a specific set of 6 global two qubit
gates can be performed, and if the initial state of the lattice can be
suitably chosen, then a quantum computer can be efficiently simulated.
\end{abstract}

\pacs{03.67.Lx}

\maketitle

\section{Introduction}

Building a computer that operates coherently at the quantum level may
revolutionise the way we carry out computations. Indeed it is believed
that such quantum computers are much more powerful than their
classical analogues. For instance it seems that factoring can be
carried out exponentially faster on a quantum computer than on a
classical computer. For this reason much work is being devoted to
developing physical systems in which computation can be carried out
at the quantum level.

A very attractive systems in which to implement quantum information
processing are atomic lattices. Indeed the method for realising such
lattices suggested in \cite{JBCGZ} has been
demonstrated in \cite{GMEHB}, and lattices comprising more than $10^5$
atoms have been realised. A method for carrying out interactions
between neutral atoms suggested in \cite{JBCGZ2} has been demonstrated
in \cite{MGWRHB}. This method realises a global two-qubit gate which
in a few steps can entangle all the lattice, leading to cluster
states\cite{a}. Finally coherent transport of atoms over many lattice
spacings has been demonstrated in \cite{MGWRHB2} which implies that
the global two-qubit gates can be realised between atoms located many
lattice spacings away. 

On the other hand atomic lattices are affected with a fundamental difficulty.
Namely it is very
difficult in these systems to address individually each atom in the
lattice. Rather one is limited to the global operations mentioned
above. Thus whereas atomic lattices seem well suited to carry out
simulations of translationally invariant physical systems\cite{Jane},
it is not as clear how to use them to implement a universal quantum
computer.

Here we address the question of the computational power of atomic
lattices. That is, to what extent can a quantum computer be efficiently
realised using atomic lattices?

We shall consider a perfect lattice, i.e. a
lattice with exactly one atom per site. We shall suppose
that the only gates which are available are global one-qubit gates
and global two-qubit gates. We will suppose that these gates 
can be performed perfectly. We shall also take
each atom to have an internal Hilbert space of dimension 2, i.e. a
qubit. These restrictions strongly 
limit the operations that can be carried
out and the core of our result consists of showing how to overcome
these constraint. Finally we shall suppose that the initial state of
the lattice breaks slightly the translational symmetry in a specific
way. Namely we shall suppose that all the atoms are initially in the
state $|0\rangle$ except two specific atoms that are in the state
$|1\rangle$. Our main result is to show that in this situation it is
possible to efficiently simulate a quantum computer.

We note that 
experiments so far involving atomic lattices have only used qubits (as
we do), but also have only implemented a single global two-qubit Hamiltonian
which in the notation below is $|01\rangle\langle 01|^{(d)}$. On the
other hand the result we report here requires two different global
two-qubit gates and global one-qubit gates.  
Whether or not the Hamiltonian  $|01\rangle\langle 01|^{(d)}$ and arbitrary
global one-qubit gates are enough to 
simulate a quantum computer is an open question. The results obtained
here may provide an avenue for tackling this problem. We expect they
will also  be
of interest in other contexts as they provide a non trivial way of
implementing a quantum computer in a system where limited sets of
gates are realisable.

The question of the computational power of atomic lattices 
has recently been studied in a number of
works. 
For instance the proposals of 
\cite{CDJWZ} and \cite{KP} are based on the concept of
a ``marker qubit'' which is circulated through the lattice. 
And 
\cite{VSC} uses as ingredient imperfections in the lattice.
The latter work has been extended in \cite{VC2} to perfect lattices
and translationaly invariant initial states.
 The techniques used in these works 
are very different from the ones presented
here. Essential differences concern for instance the size of the
Hilbert space of each atom, the initial state, and the way local gates
between logical qubits are implemented using the global gates.

Finally it may be interesting to note that the present work was
motivated by a numerical study of the computational power of atomic
lattices. In this numerical investigation 
we allowed all global 
 one-qubit gates and a single global two-qubit gate on qubit
pairs of distance 1 on a lattice consisting of
$n$ qubits on a circle. We considered the action
of these global gates on the eigenspaces of the cyclic
shift operator in the Hilbert space of the states
of the $n$ qubits corresponding to the eigenvalue 1. (As
the shift operator commutes with the global gates, 
all of its eigenspaces are invariant under the action
of global gates.) Using the computational algebra
system GAP \cite{GAP04}, we obtained that for
$n=3,\ldots,7$ qubits, the restriction of the
Hamiltonians of global gates to the eigenspace
generate the whole unitary Lie algebra. That is,
at least up to 7 qubits, the above global gates form
a universal set of gates on the eigenspace. If this
holds for every $n$ (as we conjecture), global
one-qubit gates and a single global two-qubit gate 
on a cyclic lattice consisting
of $n$ qubits can implement arbitrary unitaries on a Hilbert space of size
roughly $n-O(\log_2 n)$ qubits. 

A possible reason for the universality we found
is that the global two-qubit gate considered has almost as many 
eigenvalues as possible. But this means that in some
sense this gate acts "chaotically". Therefore the model
is probably not very useful in the sense that  
it does not seem to allow one to 
define a qubit structure on the
eigenspace in a natural way. 
For this reason we turned to the model described above which uses
more two-qubit gates, which allows a qubit structure to be defined, 
 and which is amenable to analytic treatment. It
is this analysis we report here.

\section{Global two-qubit gates}

We begin by giving a precise definition of global two-qubit gates.

Let $D$ be a subset of an abelian group $G$ where  
$|D|=n$. The Hilbert 
space of the pure states of the $n$ qubits is $\C^{2^n}$. 
The elements of the standard basis
are indexed by the functions $D\rightarrow\{0,1\}$.
For a function $a:D\rightarrow \{0,1\}$ the corresponding
basis element is denoted by $\ket{a}$. If $p\in D$ we also
write $a_p$ for the value $a(p)\in\{0,1\}$.

For a 2-qubit operation or 
$2^2\times 2^2$ matrix $M$
and a pair of elements $p,q\in D$, 
$M^{(p,q)}$ denotes the $n$-qubit operation which
acts as $M$ on the pair of qubits at positions $p$
and $q$:
$$M^{(p,q)}_{a,b}=
\left\{\begin{array}{ll}
M_{(a_p,a_q),(b_p,b_q)} & 
\mbox{if $a_s=b_s$ for every $s\in D\setminus\{p,q\}$,}
\\ 0 & \mbox{otherwise,} 
\end{array}\right.
$$
or, in the bra-ket notation
$$\bra{a}M^{(p,q)}\ket{b}=
\left\{\begin{array}{ll}
\bra{a_pa_q}M\ket{b_pb_q} &
\mbox{if $a_s=b_s$ for every $s\in D\setminus\{p,q\}$,}
\\ 0 & \mbox{otherwise,} 
\end{array}\right.
$$ 

We introduce a \textit{weight function} $W:D\times D\rightarrow \R$
on the pairs of $D$. This function
corresponds to the fact that the global qubit gate can act with
different strength on different pairs of atoms in the lattice. We
could take $W$ to be constant, thereby respecting the translation
invariance. 
%However it is convenient to keep it as a book-keeping device.
For a two-qubit matrix $M$ and a 
vector $d\in G$ the \textit{global operation} $M^{(d)}$ is the 
sum of all
copies of $M$ acting on pairs of qubits having difference $d$,
weighted by $W$:
$$M^{(d)}=
\sum_{\begin{array}{c}p,q\in D \\ p-q=d\end{array}}W(p,q)M^{(p,q)}.$$

A \textit{global 2-qubit Hamiltonian} is a matrix of the
form $H^{(d)}$ where $H$ is an Hermitian (i.e., self-adjoint)
2-qubit
operation and a \textit{global 2-qubit gate} is an
operation of the form $\exp({-\imagi H^{(d)}})$ where
$H^{(d)}$ is a global two-qubit Hamiltonian. 

\section{An efficient encoding}
\label{encoding-sect}

The key to our approach is to use a subset $P\subset D$ of the qubits
as logical qubits. All the qubits in $P$ should initially be in a
known state, for instance all in the state
$\ket{0}$. 
The rest of the qubits will be set initially to the
$\ket{0}$ state except for two particular qubits $r$ and $r'$ which
are set to $\ket{1}$.  The qubits in $D\setminus (P\cup
\{r,r'\})$ play a separator role in the computation and are always
brought back to $\ket{0}$ after each elementary logical gate, whereas the
qubits $r$ and $r'$ are always brought back to $\ket{1}$ after each
logical gate. The qubits $r$ and $r'$ serve as reference points in
our method. Intuitively, they are used to "locate" logical qubits 
and help to "extract" local operations at the right places from 
global ones.
The subset $P$ and $r$, $r'$
obey some geometrical constraints which we now describe.

\begin{enumerate} 
\item If $p\in P$, $s\in \{r,r'\}$, $q,q' \in P\cup \{r,r'\}$ such that
$q-q'=p-s$ then $q=p$ and $q'=s$. That is for every $p\in P$, both 
$p-r$ and $p-r'$ occur exactly once as a 
difference of a pair of points from $P\cup \{r,r'\}$. 
\item If $q,q' \in P\cup \{r,r'\}$ such that $q-q'=r-r'$ then $q=r$
  and $q'=r'$. That is $r-r'$ occurs exactly once as a 
difference of a pair of points from $P\cup \{r,r'\}$. 
\item For every $p\in P$ there exists no pair $q,q'\in P\cup\{r,r'\}$
such that $p-r+p-r'=q-q'$.
\end{enumerate}

\noindent{\bf Examples.}
It is not difficult to find groups $G$  and sets $P$, $\{r,r'\}$ that
  satisfy these constraints.

\begin{itemize}
\item An $l$ dimensional lattice of size $m$ in each direction:
$G=\Z^l$,
$D=\{0,\ldots,m-1\}^l$, 
$P=\{p=(p_1,\ldots,p_l)\in D\mid\sum_{i=1}^lp_i\equiv 0\pmod{6}\}\setminus\{0\}
$,
$r=(1,0,\ldots,0)$, $r'=(2,0,\ldots,0)$. 
Here $|P|=|D|/6-1$, i.e., roughly
every sixth element of $D$ belongs to $P$.
\item
%A one dimensional ring 
A circle of size $n=6k$:
$D=G=\Z_n$, 
$P=\{p\in G\mid p\equiv 0\pmod{6}\}\setminus\{0\}
$,
$r=1$, $r'=2$. In this example $|P|$
is again $|D|/6-1$.
\item
An $l$ dimensional lattice of size $m=3j+1$ in each direction.
$G=\Z^l$,
$D=\{0,\ldots,m-1\}^l$, %$k=\lfloor (m-1)/3\rfloor\}$,
$P=\{p=(p_1,\ldots,p_l)\in D\mid 2j+1\leq p_1\leq 3j\}$,
$r=(0,\ldots,0)$, $r'=(j,0,\ldots,0)$. Here
$|P|=jm^{l-1}=\frac{m-1}{3}m^{l-1}\approx |D|/3$.
\end{itemize}
It is not difficult to generalise these examples or combine them in
different ways.

We say that a function $a:D\rightarrow\{0,1\}^n$ is \textit{admissible}
if $a_r=a_{r'}=1$ and $a_p=0$ for every other $p\in D\setminus P$. 
Let $k=|P|$. Then functions 
$\{1,\ldots,k\}\rightarrow \{0,1\}$ can be
identified with the admissible functions in a natural way
therefore admissible functions can encode $k$ qubits.

We can now state our main theorem:

\begin{theorem}\label{TH1}
Assume that for every pair $q\neq q'\in P\cup\{r,r'\}$, we have
$W(q,q')\neq 0$. Let 
$w=\max\{|W(q,q')|\;:\;q\neq q' \in D\} /
\min\{|W(q,q')|\;:\;q\neq q'\in  P \cup \{r,r'\}\}$.
Assume further that for every pair $p\neq p'\in P$,
the following global two-qubit gates
can be implemented for any $t $ and 
$\delta\in\{0,1\}$:
\begin{equation}
\label{addressing-exp}
\exp\left(-t\imagi(\ket{11}\bra{11})^{(p-r)}\right),\;\;
\exp\left(-t\imagi(\ket{11}\bra{11})^{(p-r')}\right),
\end{equation}
\begin{equation}
\label{gates-exp}
\exp\left(-t(\ket{1\delta}\bra{0\delta}
-\ket{0\delta}\bra{1\delta})^{(p-p')}\right),\;\;
\exp\left(-t\imagi(\ket{1\delta}\bra{0\delta}
+\ket{0\delta}\bra{1\delta})^{(p-p')}\right).
\end{equation}
Then on the Hilbert space of the admissible functions
$|P|$-qubit quantum computations can be efficiently
simulated using global gates of type
(\ref{addressing-exp}) and (\ref{gates-exp}). 
Here by efficiency we mean that the complexity
of the simulation, measured in the number of
global two-qubit gates used, 
is polynomial in $w,n$ and the complexity of the original
computation.
\end{theorem}

An upper bound on the efficiency, i.e. on the degree of the
polynomial, can be obtained from
the proofs of theorems \ref{TH1} and \ref{TH6}. 
This upper bound is
probably far from optimal.

Section \ref{Proof} is devoted to the proof of this result.
However we shall first show that one can achieve the same result as
stated in Theorem  \ref{TH1} by using global one-qubit gates and fewer
global two qubit-gates.

\section{Using fewer global two-qubit gates}

The gates eq. (\ref{addressing-exp}) are the global controlled phase gates.
It is interesting to note that the global gates eq. (\ref{gates-exp})
can be thought of as generating global 
Controlled-NOT gates.  Indeed  the two-qubit Hamiltonians appearing in
these gates are
$$
\sigma_x\otimes \ket{\delta}\bra{\delta}\quad , \quad 
\sigma_y \otimes \ket{\delta}\bra{\delta}\ .
$$
which exponentiated for time $\pi/2$ yield
$$
\exp\left(-\imagi\frac{\pi}{2}
(\ket{1\delta}\bra{0\delta}
+\ket{0\delta}\bra{1\delta})\right)
= \left(-\imagi( \ket{1}\bra{0}
+\ket{0}\bra{1}) \otimes \ket{\delta}\bra{\delta} \right) 
\ ;$$
and similarly for the other gates in 
 eq. (\ref{gates-exp}),
 but in different bases. Note however that the interpretation as a
 C-NOT is not valid for the global gate, because the Hamiltonians
 acting on the different pairs of qubits do not commute.

Let us now show that if one can realise global one-qubit gates, then
the four global two-qubits eq. (\ref{gates-exp}) can all be
implemented once a single one can be implemented.
To see this we will denote a global one-qubit gate as
$$
u^{global} = \prod_{p\in D} u^{(p)}
$$
where $u^{(p)}$ is the unitary transformation that acts as $u$ on the qubit at
position $p$ only:
$$
\langle a | u^{(p)} | b\rangle = 
\left\{\begin{array}{ll}
\bra{a_p}u\ket{b_p} &
\mbox{if $a_s=b_s$ for every $s\in D\setminus\{p\}$,}
\\ 0 & \mbox{otherwise.} 
\end{array}\right.
$$ 
We then have
\begin{eqnarray}
u^{global} e^{-\imagi H^{(d)}} u^{global\dagger}
&=& e^{-\imagi u^{global}  H^{(d)} u^{global\dagger}}\nonumber\\
&=& \exp\left[ -\imagi \sum_{\begin{array}{c}p,q\in D \\ p-q=d\end{array}}
W(p,q)
u^{(p)} u^{(q)} H^{(p,q)} u^{(p)\dagger} u^{(q)\dagger}\right]
\label{u2u}
\end{eqnarray}
Using this expression it is easy to see that  
the four global 2 qubit-gates appearing eq. (\ref{gates-exp}) are equivalent
if one can implement the global one-qubit gates $\sigma_x^{global}$
and
$(1/\sqrt{2} + i \sigma_z/\sqrt{2})^{global}$.

Note that the class of global two-qubit operations that can be
implemented when a single global two-qubit operation and arbitrary global
one-qubit operations can be implemented is larger than the class given
in eq. (\ref{u2u}), see \cite{MVL}.

\section{Proof of main theorem}\label{Proof}

Our aim is to show how local gates between two qubits $p,p'\in P$ can
be efficiently implemented by sequences of global gates
(\ref{addressing-exp}) and (\ref{gates-exp}). We will first study how
this can be done at the level of Hamiltonians by commuting 
$A^{(p-r)}=
%\imagi
(\ket{11}\bra{11})^{(p-r)}$,
$A^{(p-r')}=
%\imagi
(\ket{11}\bra{11})^{(p-r')}$ 
%(that is, the Hamiltonians of the global gates 
%in (\ref{addressing-exp}))
%and a specific set of 
%global 2-qubit operations
%$B^{(p-p')}_\delta$ (which 
%are the basic constituents
%of the Hamiltonians of the global gates in (\ref{gates-exp}),
%see the definition later on). 
and the Hamiltonians of the global gates in (\ref{gates-exp}).
These results on
commutation of Hamiltonians will then imply the results for the
implementation of two qubit gates, ie. for unitary operations.

For $a,b:D\rightarrow\{0,1\}$ the elementary matrix with zeros at every
position except for $a,b$ where the entry is one is denoted
by $E_{a,b}$:
$$E_{a,b}=\ket{a}\bra{b}.$$
For $a:D\rightarrow \{0,1\}$ and $p\in D$ we denote
by $\del_p(a)$ the function that can be obtained
by zeroing the bit of $a$ at position $p$:
$$\del_p(a)_q=
\left\{\begin{array}{ll}
a_q & \mbox{if $q\neq p$,}\\
0 & \mbox{if $q=p$.}
\end{array}\right.
$$
Our first step will be to investigate how the commutations act on the
elementary matrix $E_{a,\del_p(a)}$.

Recall that 
$A$ is the two-qubit operation $\ket{11}\bra{11}$.
Its matrix is a diagonal matrix with entry one
at position corresponding to $\ket{11}$ and zero
elsewhere:
$$A_{a,b}=\bra{a}A\ket{b}=
\left\{
\begin{array}{ll}
1 & \mbox{if $a=b=11$,}\\
0 & \mbox{otherwise.}
\end{array}\right.
$$
Thus if $0\neq d\in G$ then $A^{(d)}$ is the diagonal matrix
where the element at the position corresponding to 
$a$ is just the sum of the weights of the $1-1$ pairs in $a$ having
difference $d$:
$$
A^{(d)}_{a,a}=
\bra{a}A^{(d)}\ket{a}=
\sum_{\begin{array}{c}p,q\in D\\p-q=d\end{array}}W(p,q)
\bra{a}A^{(p,q)}\ket{a}
=
\sum_{\begin{array}{c}p,q\in D \\ p-q=d \\ a_p=a_q=1\end{array}}
W(p,q).
$$

As a consequence, if $a$ is a function from
$D$ to $\{0,1\}$ and $q\in D$ with $a_q=1$, then we have the
following formula:
%elementary commutator:
\begin{equation}
\label{addr-comm-eq}
[A^{(d)},E_{a,\del_q(a)}] = 
(a_{q-d}W(q,q-d)+a_{q+d}W(q+d,q))
E_{a,\del_q(a)}. 
\end{equation}
(Here, in order to simplify notation, we assume that
$W(l,l')=0$ if $l\not\in D$ or $l'\not\in D$).
Indeed, 
\begin{eqnarray*}
[A^{(d)},E_{a,\del_q(a)}] & = &
A^{(d)}\ket{a}\bra{\del_q(a)}-
\ket{a}\bra{\del_q(a)}A^{(d)}\\
&=&
\sum_{\begin{array}{c}p',q'\in D \\ 
p'-q'=d \\ a_{p'}=a_{q'}=1\end{array}}
W(p',q')\ket{a}\bra{\del_q(a)}
\quad -
\sum_{\begin{array}{c}p',q'\in D \\ 
p'-q'=d \\ \del_q(a)_{p'}=\del_q(a)_{q'}=1\end{array}}
W(p',q')\ket{a}\bra{\del_q(a)}.
\end{eqnarray*}
All the terms of the second sum appear also in the first
one and the possible terms of the first sum missing from the
second one are 
$W(q,q-d)\ket{a}\bra{\del_q(a)}$ (if $a_{q-d}=1$)
and $W(q+d,q)\ket{a}\bra{\del_q(a)}$ (if $a_{q+d}=1$).
This gives (\ref{addr-comm-eq}).

A consequence of (\ref{addr-comm-eq}) is that 
if $a$ is admissible, $a_q=1$, $p\in P$ and
$s\in\{r,r'\}$ then
\begin{equation}
\label{comm-adm-eq}
[A^{(p-s)},E_{a,\del_q(a)}]=
\left\{\begin{array}{ll}
W(p,s)E_{a,\del_q(a)} &\mbox{if $q\in \{p,s\}$ and $a_p=1$,}\\
0 & \mbox{otherwise.}
\end{array}\right.
\end{equation}
Indeed, $q\in P\cup\{r,r'\}$ by admissibility of $a$.
If $a_{q-(p-s)}=1$ then $q-(p-s)\in P\cup\{r,r'\}$
which implies $q=p$ and $q-(p-s)=s$ by the first
constraint on $P \cup \{r,r'\}$. In addition 
$q+p-s\not\in P\cup\{r,r'\}$ and therefore
$a_{q+p-s}=0$ and the coefficient given
in (\ref{addr-comm-eq})
is $W(p,s)$. Similarly, $a_{q+{p-s}}=1$ is possible
if and only if $q=s$ and $p=q+p-s$ and in this case, using once more
the first constraint on $P\cup\{r,r'\}$
the coefficient is again $W(p,s)$
since $a_{p-(p-s)}W(p,p-(p-s))=a_sW(p,s)=W(p,s)$.
This discussion proves (\ref{comm-adm-eq}). 

On the other hand, if $a$ is not admissible but
$\del_q(a)$ is admissible then
\begin{equation}
\label{comm-unadm-eq}
[A^{(p-s)},E_{a,\del_q(a)}]=0,
\mbox{~provided that $\{q-p+s,q+p-s\}\cap (P\cup \{r,r'\})=\emptyset$.}
\end{equation}
This follows from (\ref{addr-comm-eq}) and the fact that all the
possible positions where $\del_q(a)$ can be 1 fall in the set
$P\cup\{r,r'\}$ (by admissibility of $\del_q(a)$). 

These results are the basic ingredients for proving:
\begin{lemma}
\label{addressing-lm}
Assume that $a$ is an admissible function with $a_q=1$ 
and 
$b=\del_q(a)$. Then for every $p\in P$,
we have
$$
[A^{(p-r)},[A^{(p-r')},E_{a,\del_q(a)}]]=
%[A^{(p-r')}[A^{(p-r)},E_{a,\del_q(a)}]]=
\left\{\begin{array}{ll}
W(p,r)W(p,r')E_{a,\del_q(a)} &\mbox{if $q=p$ and $a_p=1$,}\\
0 & \mbox{otherwise.}
\end{array}\right.
$$
If $a$ is not admissible but $\del_q(a)$ is admissible
then for every $p\in P$,
$$
[A^{(p-r)},[A^{(p-r')},E_{a,\del_q(a)}]]=0.
$$
\end{lemma}

\begin{proof}{Proof}
Assume that $a$ is admissible. Then repeated applications
of (\ref{comm-adm-eq}) (first for $s=r'$ and then
for $s=r$) give that
$[A^{(p-r)},[A^{(p-r')},E_{a,\del_q(a)}]]$ can
be nonzero only if $q\in \{p,r'\}\cap\{p,r\}=\{p\}$,
that is $q=p$. On the other hand, again using 
(\ref{comm-adm-eq}) twice, 
it is straightforward to verify that
$[A^{(p-r)},[A^{(p-r')},E_{a,\del_p(a)}]]=
W(p,r)W(p,r')E_{a,\del_p(a)}$. This finishes the proof
of the first assertion.

To see the second statement,
assume that the commutator is nonzero. Then, by 
(\ref{comm-unadm-eq}), 
$\{q-p+r',q+p-r'\}\cap(P\cup\{r,r'\})\neq\emptyset$
and
$\{q-p+r,q+p-r\}\cap(P\cup\{r,r'\})\neq\emptyset$.
Assume first that $q-p+r',q-p+r\in P\cup\{r,r'\}$. 
Then, using
$r-r'=(q-p+r)-(q-p+r')$ and the second constraint on $P\cup\{r,r'\}$,
we have $r=q-p+r$, $r'=q-p+r'$ and
$q=p\in P$. But then if
$a$ is not admissible then $\del_q(a)$ is not
admissible either, a contradiction. 
(The case $q+p-r',q+p-r\in P\cup\{r,r'\}$ 
can be treated in a similar way). 

Finally, assume that $q+p-r',q-p+r\in P\cup\{r,r'\}$
(the remaining case can be treated by a symmetric argument).
Then $(q+p-r')-(q-p+r)=(p-r)+(p-r')$, which is impossible by the third
property of the configuration $P\cup\{r,r'\}$.
\end{proof}

We will now use lemma \ref{addressing-lm} to show how commutations of
certain global two-qubit operators $B^{(p-q)}_\delta$ with $A^{(p-r)}$ and
$A^{(p-r')}$ yields a local two qubit operator. The operators
$B^{(p-q)}_\delta$ will be the basic constituents of the Hamiltonians
of the global gates in (\ref{gates-exp}).
We define $B_\delta=\ket{1\delta}\bra{0\delta}$ 
for $\delta\in\{0,1\}$,
i.e.  $B_0=\ket{10}\bra{00}$
and $B_1=\ket{11}\bra{01}$. 
Assume that we take an order of the basis where the first 
$2^{|P|}$ basis elements correspond to the admissible 
functions and the rest correspond to the inadmissible functions.
The next lemma states that in this order of basis
the matrix of
$[A^{(p-r)},[A^{(p-r')},B_\delta^{(p-q)}]]$
for $p,q\in P$ is block diagonal where
the upper left $2^{|P|}\times 2^{|P|}$ 
block is a scalar multiple
of the corresponding block of $B_\delta^{(p,q)}$.

\begin{lemma}
\label{comm-simu-lm}
Let $p\neq q\in P$, $a,b:D\rightarrow\{0,1\}$
such that either $a$ or $b$ is admissible. Then
$$\bra{a}[A^{(p-r)},[A^{(p-r')},B_\delta^{(p-q)}]]\ket{b}
=W(p,q)W(p,r)W(p,r')\bra{a}B_\delta^{(p,q)}\ket{b}.$$
In particular if only one of $a$ and $b$ is admissible then
$$\bra{a}[A^{(p-r)},[A^{(p-r')},B_\delta^{(p-q)}]]\ket{b}
=0.$$
\end{lemma}

\begin{proof}{Proof}
For every $a,b:D\rightarrow\{0,1\}$ and for every pair
$p'\neq q'\in D$ we have
$\bra{a}B_\delta^{(p',q')}\ket{b}=1$
if and only if $a_{p'}=1$, $b_{p'}=0$, $a_{q'}=b_{q'}=\delta$,
and $a_s=b_s$ for every $s\in D\setminus \{p',q'\}$. Otherwise
$\bra{a}B_\delta^{(p',q')}\ket{b}=0$. An equivalent formulation of
this is
$$
B_\delta^{(p',q')}=
\sum_{\begin{array}{c}a:D\rightarrow\{0,1\},\\ 
a_{p'}=1, a_{q'}=\delta\end{array}}
E_{a,\del_{p'}(a)}.
$$
{}From this equality we infer
\begin{eqnarray*}
B_\delta^{(p-q)} & = &
\sum_{\begin{array}{c}p',q'\in D,\\ p'-q'=p-q\end{array}}
W(p',q')B_\delta^{(p',q')}
\\
& = &
\sum_{\begin{array}{c}p',q'\in D,\\ p'-q'=p-q\end{array}}
\sum_{\begin{array}{c}a:D\rightarrow\{0,1\},\\ a_{p'}=1, 
a_{q'}=\delta\end{array}}
W(p',q')E_{a,\del_{p'}(a)}
\\
& = &
\sum_{\begin{array}{c}a:D\rightarrow\{0,1\}\end{array}}
\sum_{\begin{array}{c}p',q'\in D,\\ p'-q'=p-q
\\ a_{p'}=1, a_{q'}=\delta
\end{array}}
W(p',q')E_{a,\del_{p'}(a)}.
\end{eqnarray*}
Using the latter equality, Lemma \ref{addressing-lm} and the fact that
$[A^{(p-r)},[A^{(p-r')},E_{a,b}]]$ is always a scalar multiple
of $E_{a,b}$, we obtain
\begin{eqnarray*}
\bra{a}[A^{(p-r)},[A^{(p-r')},B^{(p-q)}]]\ket{b} =
%\\
%& &
\left\{\begin{array}{ll}
W(p,r)W(p,r')W(p,q) & \mbox{if $b=\del_p(a)$, $a_p=1$ and $a_q=\delta$,}
\\
0 & \mbox{otherwise,}
\end{array}\right.
\end{eqnarray*}
whenever either $a$ or $b$
is admissible. {}From this equality the assertions follow as
\begin{eqnarray*}
\bra{a}B_\delta^{(p,q)}\ket{b}=
\left\{\begin{array}{ll}
1 & \mbox{if $b=\del_p(a)$, $a_p=1$ and $a_q=\delta$,}
\\
0 & \mbox{otherwise.}
\end{array}\right.
\end{eqnarray*}
\end{proof}

{}From lemma \ref{comm-simu-lm} we easily derive a similar result regarding
the block structure of the matrices obtained by
commuting the Hamiltonians of the global gates
in (\ref{gates-exp}) with the Hamiltonians of
the global gates (\ref{addressing-exp}). The result
can be interpreted as stating that, restricted to the subspace
spanned of the admissible states, the commutators
coincide (up to a scalar multiple)  with the Hamiltonians
of the corresponding (local) two-qubit gates acting
on the pair of qubits at positions $p$ and $q$. 

\begin{proposition}
\label{comm-simu-prop}
For $\delta\in\{0,1\}$ let 
$U_\delta$ be any of the Hamiltonians
$-(\ket{1\delta}\bra{0\delta}-\ket{0\delta}\bra{1\delta})$
and $-\imagi(\ket{1\delta}\bra{0\delta}+
\ket{0\delta}\bra{1\delta})$. 
Let $p\neq q\in P$, $a,b:D\rightarrow\{0,1\}$
such that either $a$ or $b$ is admissible. Then
$$\bra{a}[-\imagi A^{(p-r)},[-\imagi A^{(p-r')},U_\delta^{(p-q)}]]\ket{b}
=-W(p,q)W(p,r)W(p,r')\bra{a}U_\delta^{(p,q)}\ket{b}.$$
\end{proposition}

\begin{proof}{Proof}
We give the proof only for 
$U_\delta=
-\ket{1\delta}\bra{0\delta}+\ket{0\delta}\bra{1\delta}$,
as the calculations for the other case
are essentially the same.
Observe that $U_\delta=-B_\delta+B_\delta^\T$. Hence,
using also that the matrices $A^{(p-r)}$ and $A^{(p-r')}$ 
are self-adjoint,
\begin{eqnarray*}
[-\imagi A^{(p-r)},[-\imagi A^{(p-r')},U_\delta^{(p-q)}]]
& = &
-[A^{(p-r)},[A^{(p-r')},U_\delta^{(p-q)}]]
\\
& = &
[A^{(p-r)},[A^{(p-r')},B_\delta^{(p-q)}]]
-[A^{(p-r)},[A^{(p-r')},{B_\delta^{(p-q)}}^\T]].
%\\
%& = &
%[A^{(p-r)},[A^{(p-r')},B_\delta^{(p-q)}]]
%-[{A^{(p-r)}}^\T,[{A^{(p-r')}}^\T,{B_\delta^{(p-q)}}^\T]]
\\
& = &
[A^{(p-r)},[A^{(p-r')},B_\delta^{(p-q)}]]
-[{A^{(p-r)}},[{A^{(p-r')}},{B_\delta^{(p-q)}}]]^\T.
\end{eqnarray*}
By Lemma~\ref{comm-simu-lm}, this gives
\begin{eqnarray*}
\bra{a}[-\imagi A^{(p-r)},[-\imagi A^{(p-r')},U_\delta^{(p-q)}]]\ket{b}
%\\
& = &
W(p,q)W(p,r)W(p,r')\bra{a}B_\delta^{(p,q)}\ket{b}
-
W(p,q)W(p,r)W(p,r')\bra{a}{B_\delta^{(p,q)}}^\T\ket{b}
\\
&=&-W(p,q)W(p,r)W(p,r')\bra{a}U^{(p,q)}\ket{b},
\end{eqnarray*}
whenever either $a$ or $b$ is admissible.
\end{proof}

This result can be used to show that 
local
gates on pairs of qubits in $P$ can be efficiently simulated using
global gates. To prove this we
will need some standard facts regarding approximations
of unitary operators.

For an operator $U$ on the Hilbert space
$\C^n$ we denote by $\norm{U}$ the operator
norm of $U$: $\norm{U}=\sup_{|x|=1}|Ux|$.
Note that $\norm{AB}\leq \norm{A}\cdot\norm{B}$.
If 
$\norm{A_1},\norm{A_2},\norm{B_1},\norm{B_2}\leq 1$
%(in particular, if $A_i$ and $B_i$ are unitary operators),
then we have
%$\norm{A_1A_2-B_1B_2}\leq\norm{A_1-B_1}+\norm{A_2-B_2}$.
%Indeed, 
$$\norm{A_1A_2-B_1B_2}=
\norm{(A_1-B_1)A_2+B_1(A_2-B_2)}\leq
\norm{A_1-B_1}\cdot\norm{A_2}+\norm{B_1}\cdot\norm{A_2-B_2}
\leq
\norm{A_1-B_1}+\norm{A_2-B_2}.$$
By an easy induction we obtain
\begin{equation}
\label{error-add}
\norm{A_1\cdots A_N-B_1\cdots B_N}
\leq \sum_{j=1}^N\norm{A_j-B_j},
\end{equation}
whenever $A_1,\ldots,A_N$ and
$B_1,\ldots,B_N$ are sequences of unitary operators.

\begin{lemma}
\label{comm-approx-lm}
There is an absolute constant $c>0$, such that
$$
\norm{
\left(
\exp(-\frac{\imagi}{\sqrt{N}}U^{-1})\cdot
\exp(-\frac{\imagi}{\sqrt{N}}V^{-1})\cdot
\exp(-\frac{\imagi}{\sqrt{N}}U)\cdot
\exp(-\frac{\imagi}{\sqrt{N}}V)
\right)^{N}-
\exp([-\imagi U,-\imagi V])
}<
c\cdot M^3N^{-\frac{1}{2}}
$$
for any $N>M^2$, where
$U$ and $V$ are Hermitian operators on the
Hilbert space $\C^n$ and $M=\max\{\norm{U},\norm{V},1\}$. 
\end{lemma}

\begin{proof}{Proof}
We use the first three terms in the expansion of
$\exp(-t\imagi U)$:
$$\exp(-t\imagi U)=1-t\imagi U-\frac{1}{2}t^2U^2+O(M^3t^3),$$
as $t\rightarrow 0$. The norm of the error term 
can be indeed upper bounded by
$$\sum_{j=3}^\infty\frac{1}{j!} (tM)^j
=M^3t^3\sum_{j=0}^\infty\frac{1}{(j+3)!}(tM)^j
<M^3t^3\sum_{j=0}^\infty\frac{1}{j!}(tM)^j
=M^3t^3e^{tM}
\leq e\cdot M^3t^3$$
if $t<1/M$. Doing the same for $\exp(-t\imagi V)$,
$\exp(t\imagi U)$, and $\exp(t\imagi V)$ and
collecting the terms with exponent greater then 2
we obtain
\begin{eqnarray*}
&
\exp(-{\imagi}{t}U)^{-1}\cdot
\exp(-{\imagi}{t}V)^{-1}\cdot
\exp(-{\imagi}{t}U)\cdot
\exp(-{\imagi}{t}V)
&\\&
=(1+\imagi tU-\frac{t^2}{2}U^2) 
(1+\imagi tV-\frac{t^2}{2}V^2)
(1-\imagi tU-\frac{t^2}{2}U^2) 
(1-\imagi tV-\frac{t^2}{2}V^2)
+O(M^3t^3)
&\\&
=1+t^2(VU-UV)+O(M^3t^3)=
1+t^2[\imagi U,\imagi V]+O(M^3t^3).
\end{eqnarray*}
On the other hand, taking just the first two
term of the expansion of $\exp([-t\imagi U,-t\imagi V])$,
we obtain
$\exp([-t\imagi U,-t\imagi V])\approx
1+t^2[\imagi U,\imagi V],
$
where the norm of the error term can be
upper bounded by
$$
(2M^2)^2t^4e^{4M^2t^2}\leq
4e^4\cdot M^4t^4\leq 4e^4M^3t^3
$$
whenever $t<1/M$. This, and the preceding formula
gives
$$
\norm{
\exp(-{\imagi}{t}U^{-1})\cdot
\exp(-{\imagi}{t}V^{-1})\cdot
\exp(-{\imagi}{t}U)\cdot
\exp(-{\imagi}{t}V)
-
\exp[-t\imagi U,\exp -t\imagi V]
}\leq c\cdot M^3t^3
$$
for $t<1/M$ with some constant $c$. Writing $t=1/\sqrt N$ in the latter
inequality we obtain the asserted result using formula
(\ref{error-add}).
\end{proof}

Now we are in a position to prove our main technical
result from which theorem \ref{TH1} will easily follow.

\begin{theorem}\label{TH6}
\label{comm-simu-thm}
Assume that for every $p'\neq q'\in D$ and $p''\neq q''\in P\cup\{r,r'\}$ we
have $\frac{|W(p',q')|}{|W(p'',q'')|}\leq w$. Then, for every
real $-1\leq T \leq 1$, for every $p\neq q\in P$, for every
$\delta\in\{0,1\}$, and for every $0<\epsilon<1$,
operations which act as
the two qubit gates 
\begin{eqnarray}
\label{local-gates-exp}
\exp\left(-T(\ket{1\delta}\bra{0\delta}
-\ket{0\delta}\bra{1\delta})^{(p,q)}\right),\;\;
\exp\left(-T\imagi(\ket{1\delta}\bra{0\delta}
+\ket{0\delta}\bra{1\delta})^{(p,q)}\right)
\end{eqnarray}
on admissible states
can be $\epsilon$-approximated
by a product of $(nw/\epsilon)^{O(1)}$
global gates of type
(\ref{addressing-exp}) and (\ref{gates-exp}).
\end{theorem}

\begin{proof}{Proof}
Let $U$ stand for any of the global Hamiltonians
$\imagi(\ket{1\delta}\bra{0\delta}
-\ket{0\delta}\bra{1\delta})^{(p-q)}$
and
$
(\ket{1\delta}\bra{0\delta}
+\ket{0\delta}\bra{1\delta})^{(p-q)}.
$
Let $V=[1/W(p,r')\ket{11}\bra{11}^{(p-r')},-\imagi T/W(p,q)U]$.
Note that for any $p,q\in P$, for any $s\in \{r,r'\}$,
$\norm {-\imagi T/W(p,q)U} \leq 2 n w$,
$\norm{ 1/W(p,s)\ket{11}\bra{11}^{(p-s)}} \leq n w$,
$\norm{ 1/W(p,r)\ket{11}\bra{11}^{(p-r)}} \leq n w$,
$\norm {V} \leq  4 w^2 n^2$.

By Proposition~\ref{comm-simu-prop}, we need to approximate the operation
$\exp([-\imagi/W(p,r)\ket{11}\bra{11}^{(p-r)},-\imagi V])$. 
By Lemma~\ref{comm-approx-lm} 
this can be done with
error at most $\epsilon/2$ using a product of $O(N_2)$ operations
which are either global operations 
of the form $\exp(\pm \imagi/(W(p,r)\sqrt{N_2}) \ket{11}\bra{11}^{(p-r)})$
or operations of the form $\exp(\pm \imagi/\sqrt {N_2} V )$,
where $N_2=O(w^{12}n^{12}/\epsilon^2)$. 

Furthermore by formula (\ref{error-add}), we obtain
an $\epsilon$-approximation if we use 
$\epsilon/(2N_2)$-approximations 
instead of the operators $\exp(\pm \imagi\sqrt {N_2} V /)$.
By Lemma~\ref{comm-approx-lm} we can 
$\epsilon/(2N_2)$-approximate the operators
$\exp(\pm \imagi\sqrt{N_2} V )=
\exp(\pm[-\imagi W(p,r')^{-1} N_2^{-1/4} \ket{11}\bra{11}^{(p-r')} ,
-\imagi T W(p,q)^{-1}N_2^{-1/4}) U ])$ by a product of $N_1$
global gates where $N_1=O(w^6 n^6 N_2^{1/2} \epsilon^{-2})=
O(w^{12}n^{12}\epsilon^{-3})$. 

The total number
of global gates used in the approximation of the
"local" one is $O(N_1N_2)=O(w^{24}n^{24}\epsilon^{-5})$.
\end{proof}

\begin{lemma}
\label{univ-gates}
The gates 
\begin{equation}
\label{gates2-exp}
\exp\left(-t(\ket{1\delta}\bra{0\delta}
-\ket{0\delta}\bra{1\delta})\right),\;\;
\exp\left(-t\imagi(\ket{1\delta}\bra{0\delta}
+\ket{0\delta}\bra{1\delta})\right).
\end{equation}
for $\delta\in\{0,1\}$ and for real numbers $-1\leq t\leq 1$
form a universal set of two-qubit gates.
\end{lemma}

\begin{proof}{Proof}
We claim that the following six 
$2^2\times 2^2$ matrices generate $\mbox{su}_{2^2}$
as a Lie algebra over $\R$:
\begin{eqnarray*}
U=\ket{11}\bra{01}-\ket{01}\bra{11},\;\;
Y=\imagi\ket{11}\bra{01}+\imagi\ket{01}\bra{11},
\\
T=\ket{11}\bra{10}-\ket{10}\bra{11},\;\;
X=\imagi\ket{11}\bra{10}+\imagi\ket{10}\bra{11},
\\
V=\ket{10}\bra{00}-\ket{00}\bra{10},\;\;
Z=\imagi\ket{10}\bra{00}+\imagi\ket{00}\bra{10}.
\end{eqnarray*}
Indeed, a basis of $\mbox{su}_{2^2}$ can be obtained
as
\begin{eqnarray*}
& T,\; U,\; V,\; X,\; Y,\; Z,
& \\
& \lbrack V,T \rbrack =\ket{00}\bra{11}-\ket{11}\bra{00},\;\;
\lbrack T,Z \rbrack =\imagi\ket{00}\bra{11}+\imagi\ket{11}\bra{00},
& \\
& \lbrack T,U \rbrack =\ket{01}\bra{10}-\ket{10}\bra{01},\;\;
\lbrack Y,T \rbrack =\imagi\ket{01}\bra{10}+\imagi\ket{10}\bra{01},
& \\
& \lbrack \lbrack V,T \rbrack ,U \rbrack =\ket{00}\bra{01}-\ket{01}\bra{00},\;\;
\lbrack \lbrack T,Z \rbrack ,U \rbrack=\imagi\ket{00}\bra{01}+\imagi\ket{01}\bra{00},
& \\
& \lbrack U, Y \rbrack=2\imagi(\ket{11}\bra{11}-\ket{01}\bra{01}),
\;% & \\ &
\lbrack T, X \rbrack=2\imagi(\ket{11}\bra{11}-\ket{10}\bra{10}),
\;% & \\ &
\lbrack V, Z \rbrack=2\imagi(\ket{10}\bra{10}-\ket{00}\bra{00}).
&
\end{eqnarray*}
{}From the claim the assertion follows as the matrices $U,V,Y,$ and $Z$
are Hamiltonians of operations of the form (\ref{gates2-exp}), while
$T$ and $X$ can be obtained form $U$ and $Y$, respectively, by
exchanging the two qubits.
\end{proof}

We now prove our main theorem:

\begin{proof}{Proof of Theorem \ref{TH1}}
Consider a quantum computation (circuit) 
on $|P|$ qubits.
Because of Lemma~\ref{univ-gates},
we may assume that the circuit is
given as a product of $\ell$ gates of the form
(\ref{gates2-exp}) acting on qubit pairs in $P$,
i.e., gates given in (\ref{local-gates-exp}).
(The complexity in terms of other, more standard 
gate set is polynomially related
to $\ell$.) Let $\epsilon>0$. By (\ref{error-add}), we 
obtain an $\epsilon$-approximation of the circuit if
we use $\epsilon/\ell$-approximations of the gates.
By Theorem~\ref{comm-simu-thm}, the effect of
an individual 2-qubit gate on admissible configurations 
can be approximated with error at most $\epsilon/\ell$
using $O((w n \ell/\epsilon)^k)$ global two-qubit gates
of the form (\ref{addressing-exp}) and
(\ref{gates-exp}) for some constant $k\leq 24$. 
In view of this, simulation of the entire
circuit requires $O(\ell( w n\ell/\epsilon)^k)
\leq O((w n\ell/\epsilon)^{k+1})$ global gates.
\end{proof}

\section{Conclusion}

In the present work we have considered the computational power of 
a lattice composed of a two dimensional system
(a qubit) at each site. The only gates we used were global two-qubit
gates which act in a translationaly invariant manner on pairs of
qubits. The initial state of the lattice consists of all qubits in the
$|0\rangle$ state, except two specific 
qubits which are in the $|1\rangle$ state. 
With these ingredients we have shown that it is possible to
efficiently simulate a quantum computer. We hope these results will
stimulate further work on the computational power of lattice systems.

Preliminary investigations suggest that one can extend the present work
in several directions. First of all it should be possible to decrease the
number of different types 
of global two-qubit gates which are used in the simulation.
Secondly we have not exploited in the present work the global
one-qubit gates. Preliminary work shows that they can be used to
simplify some aspects of the simulation.

\bigskip

{\bf Acknowledgments:} 
We are grateful to Ignatio Cirac for presenting us the
model of global gates, for suggesting the question
of equivalence with conventional quantum computers
and also for his useful remarks and suggestions.
We thank Lajos R\'onyai for his helpful suggestions
regarding the proofs.
We acknowledge financial support by project
RESQ IST-2001-37559 of the IST-FET program of the EC, by the
Communaut\'e Fran{\c {c}}aise de Belgique under grant ARC00/05-251, 
by the IUAP program of the Belgian government under grant V-18
and 
by the Hungarian Scientific Research Fund (OTKA) under
grants T42706 and T42481.


\begin{thebibliography}{99}

\bibitem{JBCGZ} D. Jaksch, C. Bruder, J. I. Cirac, C. W. Gardiner,
  P. Zoller, Phys. Rev. Lett. {\bf 81} (1998) 3108

\bibitem{GMEHB} M. Greiner, O. Mandel, T. Esslinger, T. W. H\"ansch,
  I. Bloch, Nature {\bf 415} (2002) 39

\bibitem{JBCGZ2} D. Jaksch, H.-J. Briegel, J. I. Cirac, C. W. Gardiner,
  P. Zoller, Phys. Rev. Lett. {\bf 82} (1999) 1975

\bibitem{MGWRHB} O. Mandel, M. Greiner, A. Wildra, T. Rom, T. W. H\"ansch,
  I. Bloch, Nature {\bf 425} (2003) 937

\bibitem{a} R. Raussendorf and H. J. Briegel, Phys. Rev. Lett. 86, 5188 (2001).

\bibitem{MGWRHB2} O. Mandel, M. Greiner, A. Wildra, T. Rom, T. W. H\"ansch,
  I. Bloch, Phys. Rev. Lett. {\bf 91} (2003) 010407

\bibitem{Jane} E. Jan{\'e}, G. Vidal, W. D{\"{u}}r, P. Zoller,
J. I. Cirac, 
Quantum Information and Computation, Vol. 3, No. 1, 15-37 (2003)


\bibitem{CDJWZ} T. Calarco, U. Dorner, P. Julienne, C. Williams,
  P. Zoller, quant-ph/0403197.

\bibitem{KP} A. Kay, J. K. Pachos, New. J. Phys. 6 (2004) 126


\bibitem{VSC}
K. G. H. Vollbrecht, E. Solano, and J. I. Cirac,
Phys. Rev. Lett. 93, 220502 (2004)

\bibitem{VC2} K. G. H. Vollbrecht, J. I. Cirac,  {\em Reversible
  universal quantum computation within translation invariant systems},
quant-ph/0502143

%\bibitem{BCJCZ}
%H.-J. Briegel, T. Calarco, D. Jaksch, J. I. Cirac, P. Zoller, 
%Journal of Modern Optics, 47, 415 (2000). 

\bibitem{GAP04} 
The GAP Group, 
GAP -- Groups, Algorithms, and Programming,
Version 4.4; 2004. 
(http://www.gap-system.org)

\bibitem{MVL} L. Masanes, G. Vidal, J. I. Latorre, quant-ph 0202042


%\bibitem{C} I. Cirac, private communication



\end{thebibliography}
\end{document}